\documentclass{elsart}

\usepackage{graphicx}
\usepackage{amsmath}
\usepackage{amssymb}

\newcommand{\error}[2]{\genfrac{}{}{0pt}{1}{+#1}{-#2}}

\begin{document}

\begin{frontmatter}


\title{On the Short Distance Part of the QCD Anomaly Contribution
to the $\boldsymbol{b\to s \eta'}$ Amplitude}

\author[Oslo]{Jan O. Eeg\corauthref{cor}},
\ead{j.o.eeg@fys.uio.no}
\author[Zagreb]{Kre\v{s}imir Kumeri\v{c}ki},
\ead{kkumer@phy.hr}
\author[Zagreb]{Ivica Picek}
\ead{picek@phy.hr}

\corauth[cor]{Corresponding author.}
\address[Oslo]{Department of Physics, University of Oslo, N-0316 Oslo, Norway}
\address[Zagreb]{%
Department of Physics, Faculty of Science, University of Zagreb,
 P.O.B. 331, HR-10002 Zagreb, Croatia}

\begin{abstract}
In addressing the  $B \rightarrow \eta' K$ puzzle,
there has been a considerable hope in the literature
to resolve it by the QCD anomaly contribution to the
$b\to s \eta'$ amplitude.
This contribution corresponds to
the electroweak $b\to s g^* g^*$ transition
followed by the off-shell gluon fusion $g^* g^* \to \eta'$.
In the present paper we perform a
critical reassessment of this issue.
We show that for the hard virtual gluons in a loop there is a well
defined short distance  amplitude corresponding
to a remnant of the QCD anomaly. However, we find that it
cannot account for the measured amplitude.

In addition, we point out that the reduction of the
gluon fusion vertex for the off-shell gluons  is
compensated by an absence of the claimed suppression in
the electroweak vertex, and that 
some nonperturbative contributions 
related to the QCD anomaly may still be viable in explaining
the physical $B \rightarrow \eta' K$ amplitude.
\end{abstract}

\end{frontmatter}

\section{Introduction}
Recent measurements \cite{CLEO00,CLEO99,BABAR01,BELLE01,BELLE01b} of
 two-body charmless hadronic $B$ meson
decays confirm the so-called $B\to\eta' K$ puzzle. According to
the PDG average \cite{RPP02} of CLEO, BaBar and Belle measurements,
the $B\to K\eta' $ decay rate turned out to be unexpectedly large
when compared to the rate obtained within the standard effective
Hamiltonian approach using quark operators multiplied by Wilson
coefficients \cite{BuBL95}.
Namely, the enhancement of
\begin{align}
Br(B^+\to K^+ \eta') &= (7.5\pm 0.7) \cdot 10^{-5}\;, \nonumber \\
Br(B^0\to K^0 \eta') &= (5.8\error{1.4}{1.3})\cdot 10^{-5} \;,
\label{ketap}
\end{align}
when compared to the QCD-penguin dominated $B\to K\pi$ rates 
\begin{align}
B(B^+\to K^0 \pi^+) &= (1.73\error{0.27}{0.24})\cdot 10^{-5}\;, \nonumber \\
B(B^0\to K^+ \pi^-) &= (1.74 \pm 0.15)\cdot 10^{-5} \;.
\label{kpi}
\end{align}
calls for 
an additional contribution to the $K\eta'$ channel, either by performing
a more complete treatment within the
standard model (SM) or by invoking new physics beyond the standard model (BSM).
Since the properties of the $\eta'$-particle are related to the QCD
axial anomaly, it has been quite generally expected that the enhancement 
in (\ref{ketap}) is related to this anomaly.

Among various mechanisms considered to explain the $B\to K\eta'$
amplitude within the SM, the $b\to s\eta'$ transition plays a
distinguished role. There was an invitation in \cite{HoT97} to
consider it as a promising short-distance (SD) mechanism.
It pertains to the recently studied singlet penguin
mechanism  \cite{Ro02,BeN02}, involving two gluons fusing into $\eta'$.

In the present paper we focus
on the SD
contribution to $b\to s\eta'$
incorporating
the electroweak $b\to s g^* g^*$ transition
followed by the off-shell gluon fusion $g^* g^* \to \eta'$, where $g^*$
denotes a virtual gluon,
and $\eta'\simeq \eta_0$ corresponds to the flavour singlet state.
For low-energy gluons the $g^* g^* \eta'$ vertex  should  be
dominated by the QCD triangle anomaly analogously
to the $\pi^0 \to 2 \gamma$ amplitude.
However,
for (at least) one highly virtual
gluon
of momentum $q$, the
$g^* g^* \eta'$ vertex is expected to be suppressed like $1/q^2$
similarly to the effective $\gamma^{*}\gamma\pi^0$ vertex
\cite{DePPO90,PhP90,BaH93}.
This remnant of the anomaly for hard off-shell gluons we will call the
``anomaly tail'' contribution.
There has been a renewed interest in such contributions in view of the
hadronic contributions to $(g-2)$ that they induce \cite{HaK98,Sh01}.

A more complete discussion on the various mechanisms presented in the
literature for $B \rightarrow \eta'K$ \cite{AtS97,HoT97,KaP97},
including various versions of form factors for $g^* g^* \eta'$
\cite{AhKS97,DuKY97,AlCGK97,Ro02,BeN02,AlP00,KrP02} will be relegated to a
forthcoming paper.
In these proposals, as a rule, the gluon virtuality employed in the
$g^* g^* \eta'$  transition form factor lies below $m_{b}^2$.
In contrast, we consider highly virtual gluons, above the $m_{b}$ scale, 
corresponding to the mentioned anomaly tail contribution.

After presenting in Sect. 2 the electroweak $b\to s g^* g^*$ vertex,
in Sect. 3 we demonstrate
 how the virtual gluons are glued to the
``anomaly tail'' part of the $g^* g^*\eta'$ vertex  mentioned above.
 The resulting contribution is dubbed
the ``short distance anomaly'' (SDA) in what follows.
In the concluding section we discuss the meaning of our results and the
relation of our contributions to those existing in the literature.

\section{The Flavour Changing $\boldsymbol{b\to s g^* g^*}$ Transition}

\begin{figure}
\centerline{\includegraphics[scale=1.0]{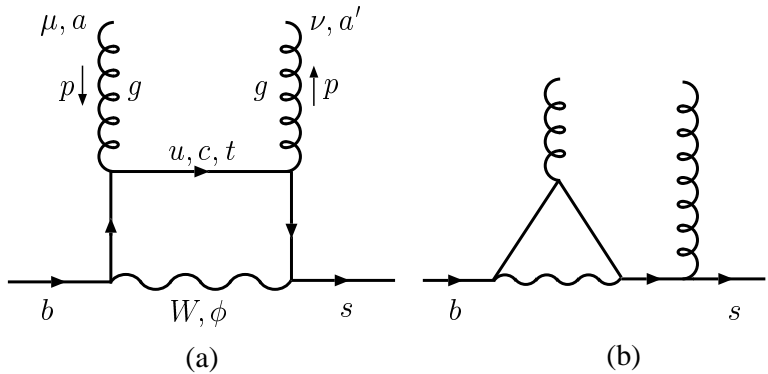}}
\caption{\label{boxtriangle}The one-loop graphs which contribute to
the box (a) and triangle (b) parts of the 
amplitude (\protect\ref{bsgg}).}
\end{figure}

The flavour changing transitions into two virtual gluons were considered
by two of the authors in the context of the double-penguin contributions
to the $K^0-\bar{K}^0$ \cite{EeP87} and $B^0-\bar{B}^0$
mixing \cite{EeP88}.
They have been subsequently studied by
Simma and Wyler \cite{SiW90} in the case of rare $B$-decays.  
Let us now reconsider these transitions by taking the symmetric
gluon momenta, so that we will be able to present the analytical expressions
suitable for evaluation of hard-gluon loop integrals. The resulting
$b\to s g^* g^*$ amplitude reads
\begin{equation}
\label{bsgg}
 M^{a a'}_{\mu \nu}\Big( b\to s \, g^*(p) \, g^*(-p) \Big) \, = \, 
\mathrm{i}\frac{\alpha_s}{\pi}\frac{G_F}{\sqrt{2}}\, \bar{s} \, t^{a'} t^a
    \sum_{i=u,c,t} \, \lambda_i H^i_{\mu \nu} \, b 
   \;+\; \text{(crossed)}\;,
\end{equation}
where $t^a$ denote colour matrices, and $\lambda_i$ are the 
Cabibbo-Kobayashi-Maskawa (CKM) factors.
$H_{\mu\nu}^i$ subsums the contribution 
 from the box 
on Fig.~\ref{boxtriangle}(a), the contribution
from the
triangle on Fig.~\ref{boxtriangle}(b), and the contribution 
from the off-diagonal $b\to s$ self-energy.
The  $b\to s \, g^*(p)$ loop in Fig.~\ref{boxtriangle}(b) is proportional to
the gluonic monopole, $\left(p^2 \gamma^{\mu} - 
p \cdot \gamma \;  p^{\mu}\right)L \, $. This $p^2$ dependence is canceled
by the $1/p^2$ dependence of the  $s$-quark propagator in this reducible 
diagram. 
After combining the triangle and the self-energy contribution, the UV
divergences mutually cancel, and only the monopole, triangle part
contributes significantly.

By anticipating the antisymmetric structure of the $\eta' g^* g^*$ vertex
into which the virtual gluons on Fig.~\ref{boxtriangle} proceed, we
select the relevant antisymmetric contribution:
\begin{equation}
\label{ti}
H^{\mu\nu}_i = (-\mathrm{i}\epsilon^{\mu\nu\rho\sigma}p_{\sigma} \gamma_{\rho}L) 
 \,A_i
  \; + \; \text{($\mu$-$\nu$ symmetric part)} \;,
\end{equation}
where we obtain
\begin{equation}\label{ai}
\begin{split}
A_i &=
-\frac{8 M_{W}^2}{m_{i}^2-M_{W}^2}\left(1+\frac{m_{i}^2}{2M_{W}^2}\right)
\int_{0}^{1} \mathrm{d}x
x(1-x)\ln\frac{D}{C} \\[2ex] 
&\quad + \frac{2 M_{W}^2}{m_{i}^2-M_{W}^2}\left(1-
\frac{m_{i}^2}{2M_{W}^2}\right) 
\int_{0}^{1} \mathrm{d}x (1-x)
 \Bigg\{ (3x-1) \mathbb{Y}_1  \\
&\qquad\qquad\qquad\qquad +
   \left[x^2(1-x)p^{2}+(x+1)m_{i}^{2}\right] \mathbb{Y}_2 
\Bigg\} \;.
\end{split}
\end{equation}
The abbreviations are
\begin{align*}
  \mathbb{Y}_1 &= 1-\frac{D}{D-C}\ln
            \frac{D}{C} \; \;  , \qquad
  \mathbb{Y}_2 = \frac{1}{D-C}\ln \frac{D}{C}
           - \frac{1}{C} \;,  \\[2ex]
 D&=x m_{i}^2 + (1-x)M_{W}^2 -x(1-x)p^2  \; , \qquad
 C= m_{i}^2-x(1-x)p^2  \; .
\end{align*}
This amplitude agrees with that of \cite{SiW90} in the region of their
mutual validity.
Let us stress that $A_i$ has the asymptotic behaviour
 $ 1/p^2$ for high gluon momenta
($-p^2\to \infty$), which 
is essential in order to obtain an overall finite gluon loop contribution
to  $b \rightarrow s \eta'$ on Fig.~\ref{gluonloop}.
Another interesting limit is the leading logarithmic approximation
\begin{equation}
\label{LLog}
  \left(  \sum_{i=u,c,t} \, \lambda_i A_i \right)_{\mathrm{L.Log}} \; = \;
\frac{4}{3} \lambda_c \ln \left(\frac{M_{W}^2}{-p^2} \right) \; ,
\end{equation}
where we have neglected the  $u$-quark contribution which is 
CKM suppressed. 
It should be noted that this leading contribution comes from the 
triangle graph in 
Fig. \ref{boxtriangle}(b), corresponding to the first term in (\ref{ai}). 
The box graph has no leading logarithm and is numerically small. 
This clearly differs from the dominance of the box part
displayed on Fig. 1 of \cite{BeN02}, 
and an explanation
on the relevance of the $b\to s g^* g^*$ amplitude (\ref{bsgg})
is in order.

There were suggestions in the literature (e. g.
\cite{KaP97}) that because of large cancellations observed by 
\cite{WiW79,LiY90,SiW90},
the contribution of the $b\to s g^* g^*$ mode followed by a $g^* g^* \to\eta'$
transition would be extremely small.
It should be strongly emphasized that this statement, valid for
soft gluons, is wrong for hard virtual gluons, as explicated in the
expressions above.
Therefore, in the next section we take under scrutiny the
amplitude stemming from the antisymmetric part of the
electroweak amplitude on Fig.~\ref{boxtriangle}.

\section{Short Distance Anomaly Contribution}

In this section, we will consider the anomaly tail part of the effective 
$g^* g^* \to \eta'$ vertex to be connected to the $b\to sg^* g^*$ amplitude 
from the preceding section. 
By restricting to the hard gluons, having a virtuality above the $m_b$
scale, we aim at singling out the short distance amplitude for the
$b\to s \eta'$ transition. 
This amplitude is of the same order in 
$\alpha_s$ as  recently studied amplitudes  with (colour)
singlet penguin topology \cite{Ro00,ChR01,Ro02,BeN02}.

\begin{figure}
\centerline{\includegraphics[scale=1.2]{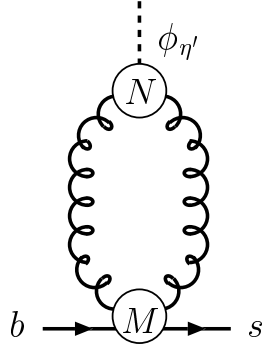}}
\caption{\label{gluonloop}The hard gluon loop contribution to the
$b \to s \eta'$ transition, determined by vertices (\protect\ref{bsgg}) 
and (\protect\ref{etagg})}
\end{figure}

\begin{figure}
\centerline{\includegraphics[scale=1.0]{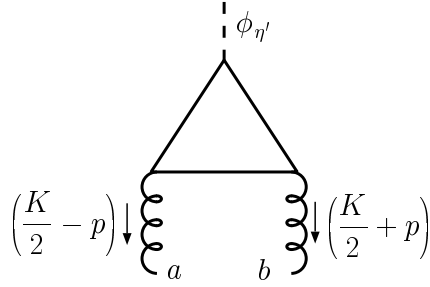}}
\caption{\label{etaanomaly}The $\eta' \to g^* g^*$ amplitude with 
indicated symmetric kinematical choice.}
\end{figure}

We need an expression for
the gluonic triangle contribution 
for $g^* g^* \to \eta'$ on
Fig.~\ref{etaanomaly}. Using the kinematical choice 
$q_1=K/2 + p$, $q_2 = K/2 - p$ for the two gluon momenta in this figure,
we get a general expression for the $g^* g^* \to \eta'$ vertex
\begin{equation}
\label{etagg}
 N^{a a'}_{\mu\nu} \Big(g^*(q_1) \, g^*(q_2) \to \eta'(K)\Big) =
  -\mathrm{i} \,\delta^{a a'}
 \epsilon_{\mu\nu\alpha\beta}\, p^\alpha K^\beta \;  G(p^2).
\end{equation}
In our case, $G(p^2)$ will turn out to be a remnant of the gluonic
anomaly.
A priori, the quantity $G$ depends also 
on the momentum $K$, but for our purposes we only need to keep
$K$ to first order there, which
 means that $G$ is taken as independent of the $\eta'$ momentum.

Our effective vertices (\ref{bsgg}) and (\ref{etagg})
allow us a perturbative evaluation of the amplitude displayed on
Fig.~\ref{gluonloop}: 
\begin{equation}
 A_{\text{SDA}}(b\to s\eta') = \int \frac{\mathrm{d}^4 p}{(2\pi)^4} \; 
 M^{a a'}_{\mu \nu} \Big( b\to s \, g^* \, g^* \Big) \;
\frac{-\mathrm{i} g^{\mu \alpha}}{p^2} 
N^{a a'}_{\alpha \beta} \Big(g^* \, g^* \to \eta'\Big)
\frac{-\mathrm{i} g^{\nu \beta}}{p^2} 
 \; . \label{SPloop}
\end{equation}
Having in mind that both vertices $M$ and $N$ imply hard gluons in
the loop, we arrive at the short distance representation of the
singlet penguin  diagram discussed in \cite{Ro00,ChR01}.
After substitution of (\ref{bsgg}) and (\ref{etagg}),
our result for the short distance di-gluon mechanism
 reads 
\begin{equation}
 A_{\text{SDA}}(b\to s\eta') =  2 i
 \, \frac{G_{\text{F}}}
{\sqrt{2}} \,(\bar{s}  K \cdot \gamma \, L \, b
) \,\sum_{i=u,c,t} \lambda_{i}
\int \frac{\mathrm{d}^4 p}{(2\pi)^4} \; 
\frac{1}{p^2}\; \frac{\alpha_{\text{s}}}{\pi}
 \; A_i(p^2) \; G(p^2) \; .
\label{asda}
\end{equation}
Since $A_i(p^2)$ and $ G(p^2)$  are already one-loop  quantities,
$A_{\text{SDA}}(b\to s\eta')$ is a 3-loop amplitude.

Let us stress that eqs. (\ref{SPloop})
and (\ref{asda}) apply quite generally for all $g^* \, g^* \, \eta'$
form factors $G(p^2)$.  In this paper we will use these formulae for the
SDA contribution only.

The evaluation of the $g^* g^* \to \eta'$ transition amplitude for
off-shell gluons is theoretically analogous to the evaluation of the 
``off-shell anomaly'' for the flavour triplet axial current related to
$\pi^0 \to \gamma^* \gamma^*$. 
A naive treatment of this photonic case
leads to problems with unitarity for the
$e^+ e^- \to \gamma^* \to \pi^0 \gamma$ amplitude presented
by Jacob and Wu \cite{JaW89}, and a 
number of papers \cite{DePPO90,PhP90,BaH93}  has been devoted
to its cure. 

The pertinent triangle amplitude was calculated first by 
Rosenberg \cite{Ro63} and
reconsidered by Adler \cite{Ad69}, who observed 
difficulties with divergences when inserting the triangle loop 
into the next loop. 
On the other hand, for off-shell photons, it was shown by
\cite{PhP90,BaH93} that the perturbative mass-independent 
part in the triangle diagram is canceled by the pion pole 
anomalous contribution. This observation was essential to resolve
 the Jacob-Wu paradox \cite{JaW89}.
In our gluonic case it means that, although the quantity $G$ in (\ref{etagg}) 
obtained in a perturbative calculation of the quark-triangle loop is
a priori proportional to 
\begin{equation}\label{ktj}
\sum_{j=u,d,s} \int_{0}^1 \mathrm{d}x \int_{0}^{1-x} \mathrm{d}y \left\{1+\frac{m_{j}^2}
{Q_j}\right\} \;,
\end{equation}
the unity term gets canceled \cite{PhP90,BaH93}.
In (\ref{ktj}) we introduced
\begin{equation}
Q_j = y(1-y) q_{1}^2 + x(1-x) q_{2}^2 + 2 x y\, q_1\cdot q_2 - m_{j}^2 
\end{equation}
representing the denominator 
emerging from the triangle-loop integral.
 
For the $g^* g^* \to \eta'$ transition the flavours $j=u,d,s$ contribute
(eventually only the $s$-quark contribution turns out to be important),
and we obtain
\begin{equation}
\label{GAn}
 G =
 \frac{1}{f_{\eta'}} \frac{\alpha_s}{\pi}
  \; \sqrt{\frac{2}{3}} \, \sum_{j=u,d,s} F_j(p, K),
\end{equation}
where $f_{\eta'} \simeq f_{\pi}\simeq 92$ MeV and
\begin{equation}
\label{FAn}
 F_j(p, K) = \int_{0}^1 \mathrm{d}x \int_{0}^{1-x} \mathrm{d}y  
 \: \frac{m_{j}^2}{Q_j}
 \; = \; - \left(\frac{m_j^2}{-p^2}\right) \; 
 \ln \left(\frac{-p^2}{m_j^2}\right) \; .
\end{equation}

\begin{figure}
\centerline{\includegraphics[scale=0.5,clip]{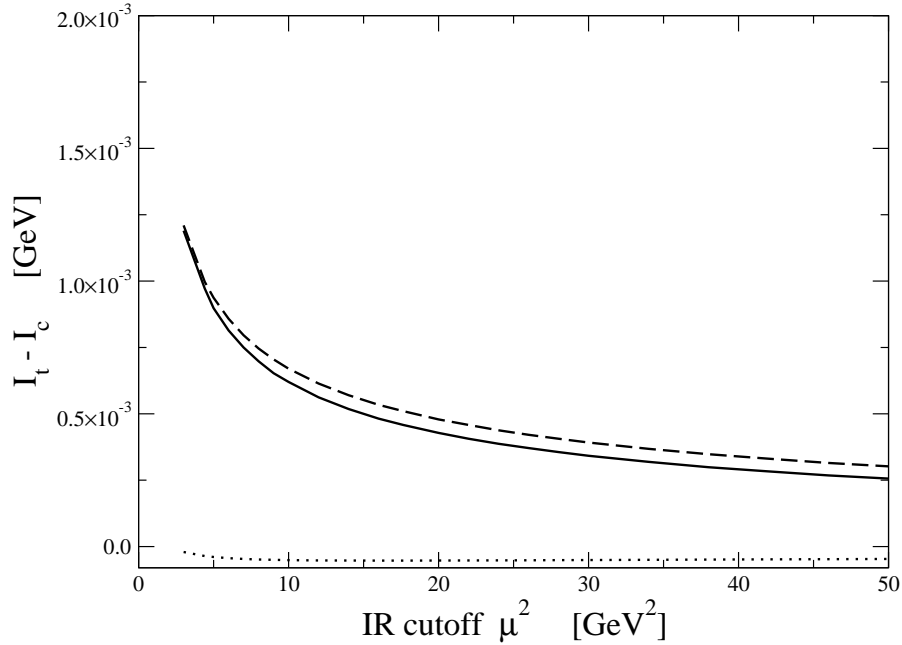}}
\caption{\label{tail} The infra-red stability check of the anomaly tail
contribution (solid line) displaying also the dominance of the triangle part
(dashed line) over the box part (dotted line) from Fig.
\protect\ref{boxtriangle}.}
\end{figure}

This represents the short distance amplitude  for hard gluon virtualities
far above the $\eta'$-mass. 
This effective vertex enters into the final gluonic
loop on Fig.~\ref{gluonloop} for loop momenta $p^2 \gg K^2 = m_{\eta'}^2$.
In the evaluation of the loop integral producing the
short distance 3-loop contribution for the $b \to s \, \eta'$ amplitude,
the overall renormalization scale $\mu$ will be 
of the order $m_b$. Taking into account that  $\mu$ acts as the effective 
IR cut-off in our loop integral, we find that the expression (\ref{FAn})
is slightly modified ($Q_j \rightarrow Q_j + \mu^2$ in (\ref{FAn}) above).
 In the leading logarithmic approximation we obtain
for the $s$-quark term
\begin{equation}
\label{etaggEff}
F_s \; = \; - \left(\frac{m_s^2}{-p^2}\right) \; \ln \left(\frac{-p^2}{\mu^2}\right) \; ,
\end{equation}
while the $u$- and $d$-quark contributions are completely negligible due
 to their small masses.
On account of the CKM unitarity, the loop integral parts
in  (\ref{asda}) 
\begin{equation}
 I_{i} = \mathrm{i} \int \frac{\mathrm{d}^4 p}{(2\pi)^4} \; 
\frac{1}{p^2}\; \frac{\alpha_{\rm s}}{\pi} \; A_i(p^2) \; G(p^2) \; ,
\label{I-integral}
\end{equation}
form the relevant GIM combination displayed on Fig.~\ref{tail}. 
This plot exhibits
a reasonably mild dependence on the infra-red cut-off ($\mu$) 
in the gluon loop,
justifying a sensible short-distance amplitude. Note that the
 UV  convergence  of the loop integral is guaranteed by the high energy 
behaviour of (\ref{ai}) and (\ref{etaggEff}).
Since the used form-factor (\ref{etaggEff}) corresponds to the
anomaly tail term, the net contribution
can be termed the  ``short distance anomaly'', as suggested in the Introduction.
Also, Fig.~\ref{tail} shows the dominance of the triangle with respect
to the electroweak box contribution. 
In the leading logarithmic approximation, we obtain from
eqs. (\ref{LLog}), (\ref{etaggEff}) and (\ref{I-integral})
\begin{equation}
 \left(\sum_{i} \lambda_i I_{i}\right)_{\rm L. Log} =\frac{\text{i}}{16 \pi^2} 
\frac{1}{f_{\eta'}} \sqrt{\frac{2}{3}} \left( \frac{\alpha_s}{\pi} \right)^2
\frac{4}{3} \lambda_c   m_s^2 
 \;  \frac{1}{6} \left\{ \ln \left( \frac{M_W^2}{\mu^2} \right) \right\}^3\ \;  \; ,
\label{I-intLL}
\end{equation}
which may be compared with the result from the (octet) penguin operator to 
lowest order in the leading logarithmic approximation 
\begin{equation}
 M_{\text{Peng}} =  \; \lambda_c
 \,  \frac{G_{\text{F}}}{\sqrt{2}} \, \left(- \frac{2}{3}  
\frac{\alpha_{\text{s}}}{\pi} \ln (\frac{M_W^2}{\mu^2}) \right)  
  (\bar{s} t^a \gamma^\mu \, L \; b) \; (\bar{q} \, t^a  \gamma_\mu \, q
)  \; \, .
\label{Peng}
\end{equation}

\section{Discussion and Conclusions}

 In the present letter we attempt to clarify a possible role of
the $b \rightarrow s\eta'$ transition in explaining the $B \rightarrow \eta' K$
amplitude, and in particular the role of the QCD anomaly in obtaining the
$b \rightarrow s\eta'$ amplitude. The result above shows that we are able to 
successfully distinguish the short distance $b\to s\eta'$ amplitude (SDA) 
related  to the QCD axial
anomaly. Thereby, the involved form of the flavour-changing vertex
for  hard gluons is in clear contrast to the result obtained by
\cite{BeN02} for soft gluons, where the box (Fig.~\ref{boxtriangle}(a))
instead of the triangle (Fig. \ref{boxtriangle}(b)) dominates. This
enhancement of the electroweak vertex for the
hard off-shell gluons is compensated by a suppression in the two gluon fusion
to $\eta'$ vertex.  The net $b \to s \eta'$
amplitude coming from this anomaly tail turns out to be dominated by the
strange quark contribution.
In the present approach it corresponds to the quark triangle contribution
to the $\eta' g^* g^*$ coupling for the highly off-shell gluons.
We note that the SDA result vanishes in the chiral limit $m_s \to 0$.

Thus, demonstrating that we have obtained 
 a  contribution which is different 
from those already existing in the literature, we can attempt to compare
it to some related amplitudes.
A simple comparison of our singlet amplitude (\ref{asda}) at the leading
logarithm level (\ref{I-intLL}) to the ordinary penguin in (\ref{Peng}):
\begin{equation}
  A_{\mathrm{Peng}} \sim f_\pi G_F 
\left(\frac{\alpha_s}{\pi}\right) \ln \left(\frac
 {M_W^2}{\mu^2} \right)
\end{equation}
gives the ratio 
\begin{equation}
\frac{A_{\text{SDA}}}{A_{\text{Peng}}} \sim 
\left(\frac{2}{3}\right)^{3/2}
\frac{\alpha_s}{\pi}
\left( \frac{m_{s}}{4 \pi f_{\eta'}} \, 
\ln \frac{M_W^2}{\mu^2} \right)^2  \; \, ,
\end{equation}
where we didn't explicate some additional factors in $A_{\rm SDA}$ and
$A_{\rm Peng}$ which cancel in the ratio.
This ratio is at the level of a few percent, but depends
strongly on what one takes for the involved parameters.
How this estimate will be modified within a more proper treatment of 
renormalization group equations is also an issue to be further 
investigated. 
We have done a simple estimate which shows that short distance
QCD corrections do not change this result substantially.

To conclude, the above demonstrates that purely short-distance
anomalous (SDA) aspects
of the $b\to s\eta'$ vertex are marginal in explaining the
$B\to \eta' K$ puzzle. There was a similar fate for the electroweak
$s \bar{d} \to \gamma^* \gamma^* \to \mu\bar{\mu}$ contribution to
the short-distance $K\to\mu\bar{\mu}$ amplitude \cite{EeKP98}.
There is an additional experience in the analogous off-shell photon 
case \cite{HaK98}, where the corresponding QED quark triangle overshoots
the measured value of the related form factor.

In our case, the smallness of the anomaly contribution to the
$b\to s\eta'$ can be ascribed to the depletion of the QCD off-shell
triangle vertex. Accordingly, the expectation of \cite{HoT97}, as we
understand it, turns
out not to be fullfiled, and we seem to be in compliance with
the conclusions by \cite{BeN02} that the singlet penguins do not
do the job and that a combination of several effects within the
Standard model would be necessary. 
In addition, our observation of the absence of the suppression in
the electroweak vertex, when we departure from truly soft gluons, opens
a window for some other
contributions which resemble the QCD anomaly. For example, ref.
\cite{Fr97} advocated that the $B\to \eta' K$ puzzle could be explained
by the additional complicated non-perturbative quark-gluon interactions
related to the anomaly. However, at present these interactions are merely 
parametrized
by a phenomenological coupling. Such an extra piece to the usual effective
Hamiltonian, appearing due to non-perturbative
aspects of the QCD anomaly, has to be justified yet.
 We intend to come back to a more complete treatment of the $B\to \eta' K$
amplitude  in a forthcoming paper. 

\begin{ack}
Two of us (K.K. and I.P.) gratefully acknowledge the support of the
Norwegian Research Council and the hospitality of the Department of
Physics in Oslo.
\end{ack}


\end{document}